\documentclass[12pt]{article}

\usepackage[top=2.72cm, bottom=2.08cm, left=2.61cm, right=2.61cm]{geometry}  
\usepackage{mathptmx}      
\usepackage[T1]{fontenc}   
\usepackage[utf8]{inputenc}
\usepackage{microtype}     
\usepackage[labelfont=bf]{caption} 

\usepackage{graphicx}      
\usepackage{booktabs}      
\usepackage{threeparttable}
\usepackage{array}         
\usepackage{float}         
\usepackage{setspace}      
\usepackage{rotating}      

\usepackage[justification=raggedright,singlelinecheck=false]{caption} 
\usepackage{xcolor}

\usepackage[none]{hyphenat}
\usepackage[english]{babel}  

\usepackage{amsfonts}      
\usepackage{amsmath}       
\usepackage{nicefrac}      

\usepackage{natbib}        
\usepackage{doi}           
\usepackage{url}           
\usepackage{hyperref}      

\usepackage{authblk}       

\usepackage[hang, flushmargin]{footmisc}
\title{The Effects of Generative AI on Collaboration and Social Networks Among Knowledge Workers}

\author[a]{Giuseppe Soda}
\author[a]{Alessandro Iorio}
\author[b]{Leonardo Rizzo}

\affil[a]{Bocconi University. Emails: giuseppe.soda@unibocconi.it; alessandro.iorio@unibocconi.it}
\affil[b]{SDA Bocconi School of Management. Email: leonardo.rizzo@unibocconi.it}

\date{}

\begin{document}
\maketitle

\begin{abstract}
\noindent \normalsize
The rapid diffusion of generative artificial intelligence has raised concerns that knowledge work may become less dependent on human-to-human interaction. Yet because AI-generated outputs require interpretation, verification, and social legitimation, generative AI may instead increase collaboration among workers. We test these competing predictions using passively collected behavioral data from more than 12,000 knowledge workers observed over 13 months during the \textcolor{black}{staged organizational rollout} of an AI-powered assistant (Microsoft Copilot). Using two complementary quasi-experimental designs---a staggered difference-in-differences model and a fixed-effects instrumental-variable approach---we estimate the effects of generative AI on workers' collaboration behavior and social network structure. We estimate that generative AI increased collaboration time across communication channels and reshaped individual networks. Those using generative AI interacted with more colleagues and their networks became less clustered, indicating broadening in collaboration structures. At the same time, they also concentrated more of their collaboration on ties with stronger prior histories, deepening reliance on already trusted colleagues. Together, these findings suggest that, while generative AI changes how people work, it may not displace human-to-human interactions.
\end{abstract}

\pagebreak
\doublespacing

\section{Main}

Cognitive work is deeply influenced by social interaction \citep{henrich2021origins}. When dealing with problems, individuals often turn to colleagues for information, judgment, interpretation, or to refine ideas until they become actionable. The social fabric of organizations is built around this dependence: who consults whom, who is consulted in return, and how these patterns of mutual reliance coalesce into networks of collaboration and trust \citep{borgatti2009network}.

Generative AI (GenAI) has the potential to alter this process in important ways.\footnote{We use the abbreviation GenAI to refer to generative artificial intelligence systems embedded in workplace tools, including Microsoft Copilot.} Earlier digital technologies primarily changed how knowledge was transmitted, stored, and accessed. For example, email, search engines, enterprise platforms, and collaboration tools accelerated communication flows and made people more productive, but they did not generate knowledge on behalf of workers. By performing such tasks as composing text, synthesizing information, and generating analyses in response to prompts, GenAI assistants become a source of cognitive output in their own right \citep{burrell2021society, licklider1960man}. Whereas earlier tools mediated connections among workers \citep{carlile2004transferring}, GenAI allows a worker to obtain answers directly from a system, reducing the need to consult another person.

This development prompts the critical question of how human-to-human collaboration changes when a non-human system performs tasks that previously required interpersonal exchange. Recent research has begun to address this issue by benchmarking human-AI collaborations against human or AI performance alone \citep{vaccaro2024when}. While these studies have advanced understanding of the consequences of using GenAI, they primarily focus on the human-machine dyad as the unit of analysis. There remains limited knowledge regarding how GenAI adoption transforms the social structure of work, specifically the patterns of interaction, dependence, and coordination among workers \citep{rastogi2023taxonomy, gonzalez2025cognitive}.

One plausible answer is relational displacement; if GenAI provides individuals with immediate, personalized access to cognitive output, it may substitute for interactions that previously required consulting peers. Workers might request a first draft from the system rather than from a colleague, seek an analysis from the system rather than from a specialist, or test an idea against the system rather than sharing it with others. In this way, generative AI may introduce a new form of technological disintermediation into human relationships, displacing not the tasks themselves, but the interactions that once surrounded them. Such changes could result in reduced collaboration time, fewer interpersonal ties, and more siloed professional networks. Similar relational patterns have been observed in research on remote work; for instance, \citet{yang2022effects} demonstrate that the shift to firm-wide remote work fragmented collaboration networks by reducing cross-group and bridging ties. Generative AI may induce a comparable form of displacement, not by eliminating physical co-presence, but by reducing the necessity to consult colleagues for information or interpretation. 

Yet this assumption remains largely untested. Research has not yet convincingly established whether GenAI actually erodes human collaboration networks. Germane to this debate is the realization that GenAI is not simply another communication technology. It produces outputs whose reliability and boundaries may be difficult for users and their peers to assess, as these systems are often characterized by invisibility, inscrutability, and constant change \citep{anthony2023collaborating}. These characteristics make the status of AI-generated outputs uncertain, including their credibility, how they should be interpreted, who should evaluate them, and when they can be shared or acted upon. Although answers may be generated rapidly, their utility depends on whether others can understand and trust them. GenAI may, therefore, introduce new coordination problems even as it reduces the cost of producing answers, shifting the bottleneck from output generation to validation and legitimation \citep{glickman2025how}. As a result, workers may become less dependent on colleagues for producing initial answers, but more dependent on them for securing acceptance. 
We argue that this dependence stems from what \citet{unceta2026epistemic} call epistemic debt. GenAI compresses the time needed to produce an output, but the time needed to understand it does not shrink accordingly. When workers accept outputs they have not fully worked through, a gap opens between what they present and what they can defend, and the gap widens as further work builds on it. \citeauthor{unceta2026epistemic} locate the remedy in individual effort: reconstructing the reasoning behind an output until one can defend it. Organizations offer a second remedy. Colleagues can question an output and ask its author to explain the reasoning behind it, which makes appropriation a social process. Put differently, if GenAI creates epistemic debt, workers may turn to peers to pay it down.
Consistent with this logic, prior research demonstrates that professionals in fields such as finance and healthcare routinely consult peers and triangulate results before taking action in response to opaque algorithmic outputs \citep{anthony2021knowledge}.

Thus, the central issue is not whether GenAI substitutes for interpersonal exchange, but whether its outputs generate new demands for validation. This ambiguity motivates the research question: Does the adoption of GenAI shrink or expand human collaboration networks within organizations? More specifically, does AI substitute for interpersonal exchange, resulting in reduced collaboration and reliance on fewer colleagues, or does it create new coordination needs, prompting increased collaboration and broader networks? Adjudicating between these competing predictions in a causal way is difficult because knowledge workers who adopt GenAI may differ systematically from those who do not in ways that also shape collaboration. Studies relying on self-reported measures, cross-sectional designs, or simple observational comparisons risk conflating sorting effects with causal effects. Our study addresses these challenges by using passively collected behavioral data from over 12,000 knowledge workers in a large corporation, observed over 13 months during the staged organizational rollout of a GenAI assistant (Microsoft Copilot) embedded in employees' daily workflows. The effects of this adoption on collaboration time and network structure are estimated using a staggered difference-in-differences design that leverages variation in rollout timing. Complementary fixed-effects instrumental-variable models further address concerns regarding the endogenous nature of GenAI usage.

The results support network expansion. Workers spend more time collaborating, engage with more peers, and show lower network clustering. But the broadening is selective. GenAI also strengthens preferential reinforcement of ties, so workers concentrate more of their current collaboration on colleagues with whom they already have strong prior histories \citep{krackhardt1992strength}. Thus, technologies that reduce the effort required to produce answers may simultaneously increase the effort required to have those answers recognized as meaningful and credible by others. In this way, GenAI does not reduce human interdependencies in the workplace but rather reorganizes them in significant ways.

\section{Methods} \label{sec:methods}

We analyze passively collected behavioral data from 12,040 employees in a large multinational organization, operating in the consumer products industry, across multiple business units and geographies. We focus on one business unit (the focal unit), which we observed continuously from September 2024 to September 2025\footnote{This project received ethical approval from the IRB of the authors' institution: Code EA001198.}. This unit implements strategic initiatives across the organization, and its work is highly interdependent and relies heavily on digital communication tools. The workforce is composed predominantly of knowledge workers whose daily activities revolve around information processing. Our observation window spans the staged rollout of the GenAI assistant, Microsoft Copilot, within this focal unit, allowing us to examine how GenAI adoption has changed collaboration and social networks over time.

All behavioral data are drawn from Microsoft Viva Insights, a workplace analytics platform that records digital collaboration activity across Microsoft 365 applications. Collaboration outcomes are measured from interactions such as calls, emails, and meetings, which capture the full range of exchanges through the digital communication channels (Teams, Outlook, and other applications). For network ties, a directed weighted tie from worker $i$ to worker $j$ is recorded in month $t$ with a weight equal to the total collaboration time that $i$ invested with $j$ across all communication channels during that period. To reduce noise from incidental interactions, we retain only ties whose weight exceeds the monthly median of all edge weights, a threshold that corresponds to approximately one hour of interaction. 

The data were anonymized at the source and restricted to knowledge workers with complete records of digital collaboration activity and Copilot license status throughout the study period. Collaboration network data are measured at the monthly level, a resolution that filters out short-term noise from episodic events and captures the more durable structural patterns of collaboration relevant to our research question. Other individual-level variables, including Copilot usage and workload indicators, were originally collected weekly. To align these with the monthly network panel, we aggregate weekly observations to the monthly level by computing the mean within each individual-month cell. For each worker, we also observe sex, age group, tenure, hierarchical level, whether they are in the HQ, and functional affiliation.

Microsoft Copilot is a GenAI assistant embedded within Microsoft 365 applications. Because it is integrated directly into the digital tools employees already use, rather than deployed as a separate technology, adoption is observable through behavioral traces rather than self-report, and the treatment is embedded in workers' natural workflow. This integration strengthens the external validity of our setting as a window into how GenAI affects everyday collaboration in knowledge work. We measure Copilot activity as the number of actions each worker performs each month. This metric captures the full range of Copilot workers' interactions available through Copilot Chat and in-app features across Microsoft 365 applications, including writing assistance, meeting summarization, document drafting, and information retrieval. An action is counted only when the user actively engages with Copilot-generated output---for example, by expanding a document summary or reviewing an automatically generated meeting recap---ensuring that the measure reflects intentional AI use rather than passive exposure.

Because our measure of Copilot activity is continuous, whereas our empirical strategy requires a binary treatment indicator, we construct a cohort-based adoption variable that captures sustained, non-trivial Copilot adoption. Specifically, a worker is coded as treated in the first month in which Copilot activity exceeds a minimum threshold, corresponding to the 25th percentile of observed use, for two consecutive months. This rule is designed to distinguish meaningful adoption from exploratory or incidental use and to satisfy the absorbing-treatment structure required by the staggered DiD estimator \citep{callaway2021difference}. From the initial population of 12,040 workers, we exclude 172 always-adopters who meet the criterion in the first observable month, since they contribute no pre-treatment observations and cannot be used to assess pre-trends. This yields a final estimation sample of 11,868 workers: 1,992 adopters, grouped into monthly adoption cohorts, and 9,876 never-adopters who serve as the control group because they do not meet the adoption criterion at any point during the observation window. 

Our primary collaboration outcome is \textit{generated collaboration hours} (i.e., generated load), a composite Viva Insights metric that captures the total meeting, call, email, and chat hours each worker initiates with at least one other person per month, after deduplication of overlapping activities (i.e., calls during a meeting). We focus on this measure because it captures active outreach---the validation-seeking behavior our argument predicts---rather than the passive receipt of communication from colleagues whose contact may be unrelated to AI-assisted work. As a robustness check, we also model total collaboration hours, which include both initiated and received collaborative activity regardless of who initiated it. Furthermore, we decompose collaboration hours by communication modality to examine whether Copilot use shifts collaboration toward more synchronous or asynchronous forms of interaction. Two clarifications about what generated load does and does not capture are worth noting. First, deduplication makes it conservative: a call occurring during a meeting is counted once, not twice. Second, "generated" reflects a specific role: the focal worker is the one who created the occasion for interaction, for example, by calling a meeting, setting its timing, or initiating an email exchange. The measure, therefore, captures the worker's role as a convener of collaboration, not their participation in interactions convened by others.

Importantly, collaboration outcomes and Copilot usage are recorded by entirely separate systems within the Viva Insights platform. While collaboration metrics are derived from Microsoft 365 activity logs (e.g., calendar events, emails, and Teams activity), Copilot usage is tracked through a dedicated logging system that records interactions with the Copilot product specifically. Because the two measures are independent by design, any association between them reflects a genuine behavioral relationship rather than a shared data source or measurement artifact.

We adopt a parsimonious, theory-driven framework centered on three dimensions of change \citep{feld2007describing} by computing three network measures from the monthly collaboration graphs to capture distinct dimensions of each worker's structural position.  
For each month $t$, we construct a directed, weighted collaboration graph $G_t = (V, E_t, w_t)$, where $V$ is the set of workers, $E_t \subseteq V \times V$ is the set of directed ties active in month $t$, and $w_t: E_t \rightarrow \mathbb{R}_{>0}$ assigns to each tie its weight. A directed tie $(i, j) \in E_t$ exists when worker $i$ generates at least one hour of collaboration with worker $j$ during month $t$, and $w_t(i, j)$ equals the total generated collaboration hours from $i$ to $j$ in that month. The one-hour threshold ensures that ties reflect substantive collaboration rather than incidental contact.\footnote{All network measures were computed in Python on the authors' University HPC cluster (Jupiter I), with each monthly network processed as an independent job. The cluster comprises 22 compute nodes with Intel Xeon processors (12--20 cores), 512--1024 GB RAM, and NVIDIA H100, H200, and A100 GPUs. Each monthly job was allocated 32 GB of memory and 4 CPU cores.} \textit{Out-degree} captures workers' active outreach by counting the number of distinct individuals to whom they direct collaboration in a given month. \textit{Clustering coefficient} is computed on the undirected projection of the collaboration network, where a symmetric tie between $i$ and $j$ exists if at least one directed tie is present. For each worker, the clustering coefficient measures the fraction of their contacts who are also connected to one another, capturing the extent to which a worker's collaborators form a cohesive, closed group. \textit{Preferential reinforcement} captures whether or not a worker concentrates current collaboration time on historically strong ties. For each worker $i$ in month $t$, we first compute the cumulative past weight of each outgoing tie $j$ as the sum of collaboration time invested across all prior months. We then compute the Pearson correlation between current tie weights and cumulative past weights across all of $i$'s outgoing ties at time $t$. Positive values indicate that the worker disproportionately allocates current collaboration time to ties in which they have invested heavily in the past; values near zero indicate no systematic relationship between current and historical tie strength. 

We estimate treatment effects using the \citet{callaway2021difference} estimator for staggered adoption designs. In essence, the design compares changes in collaboration and network outcomes following Copilot adoption with changes among never-adopters over the same period, while accounting for common time shocks and stable pre-treatment differences across workers. We implement the estimator with the doubly robust inverse-probability weighting approach, which remains consistent if either the outcome model or the treatment model is correctly specified, providing greater protection against model misspecification than either alone. Pre-treatment covariates, including sex, age group, tenure, office location, and hierarchical level, are included because they are predetermined worker characteristics that may predict both adoption timing and baseline collaboration patterns, so that parallel trends need only hold conditional on these characteristics rather than unconditionally. Standard errors are clustered at the worker level. Given our 13-month panel, we use an event-study window of $[-3,+2]$, which preserves a meaningful check for pre-treatment differences while avoiding reliance on long leads and lags identified only from the earliest or latest adoption cohorts.

The key identifying assumption is conditional parallel trends: absent Copilot adoption, treated and never-treated workers would have followed similar collaboration and network trajectories, conditional on the covariates above. The \citet{callaway2021difference} framework does not require that treatment timing be randomly assigned or exogenous; it permits selection into adoption timing on observed characteristics, provided parallel trends hold conditional on those characteristics, together with no anticipation of treatment. This assumption is relevant in our setting, where Copilot access followed organizational processes such as budget availability rather than individual initiative, but where the precise timing of sustained use remains a behavioral choice. 

As a complementary analysis, we estimate fixed-effects instrumental-variable models that use the intensity of Copilot activity rather than the binary adoption indicator. Whereas the staggered difference-in-differences estimates identify the effect of adoption, the instrumental-variable estimates isolate the intensive margin: variation in how much workers use Copilot conditional on having access. These analyses address the concern that, even after accounting for worker and month fixed effects, the amount of Copilot use may be endogenous to time-varying characteristics such as project assignments, task demands, or coordination needs. We instrument monthly Copilot activity using a shift-share design. The shift is the monthly intensity of Microsoft Copilot feature releases, constructed from Microsoft's official release notes, capturing organization-wide variation in the salience and usefulness of Copilot over time. The share is each worker's baseline exposure to Copilot users, measured as the share of the focal worker's collaborators who were already active Copilot users in that worker's first observed month in the panel. Because workers enter the panel at different times, this baseline month is worker-specific rather than a common calendar month. This captures how embedded the worker was in a subnetwork of early adopters before subsequent changes in Copilot activity. The instrument is the interaction between these two components. The identifying variation, therefore, comes from a common shock whose effective intensity differs across workers depending on their pre-existing exposure to Copilot users: Microsoft feature releases affected the entire organization, but they should have increased Copilot use more strongly among workers whose collaboration networks already contained more active users.

The identifying assumption is that, conditional on worker and month fixed effects, the shift-share instrument affects the focal worker's collaboration outcomes only through the focal worker's own Copilot use. We assess the plausibility of this assumption in two ways. On the one hand, the way we define the outcome limits one important alternative channel: because generated load captures only interactions initiated by the focal worker, collaborators' use of Copilot cannot directly increase the outcome through colleague-initiated outreach. On the other hand, we conduct two complementary diagnostic tests, following the exposure-design logic articulated by \citet{goldsmith2020bartik}, to assess whether the instrument predicts collaboration outcomes through channels other than focal-worker Copilot use. Both tests use the never-adopter sample in which Copilot use cannot mediate the relationship between the instrument and collaboration. First, we estimate a placebo reduced-form specification that regresses generated collaboration hours on the shift-share instrument, absorbing worker and month fixed effects, and clustering standard errors at the worker level. The estimated coefficient is small and statistically insignificant [$\beta = 0.011$, $SE = 0.010$, $p = 0.237$; 95\%CI ($-0.008, 0.030$)], providing no evidence that the instrument directly predicts collaboration among workers who never adopt Copilot. Second, we examine whether the exposure component of the instrument (contacts of the focal worker using AI) is associated with differential collaboration trajectories in the absence of adoption. To do so, we estimate a flexible specification in the never-adopter sample that interacts baseline exposure with month indicators, again absorbing worker and month fixed effects and clustering standard errors at the worker level. The exposure-by-month terms are jointly insignificant ($F(12, 9293) = 0.94$, $p = 0.504$), indicating no systematic evidence that baseline exposure predicts different patterns of collaboration among workers who never use Copilot. Together, these tests reduce the concern that the instrument captures peer-induced collaboration, social exposure to early Copilot users, update-related activity, or pre-existing differences in collaboration dynamics rather than variation in the focal worker's own Copilot use. Although they cannot prove the exclusion restriction, they provide additional support for the validity of the shift-share design.

\section{Results}

\subsection{Effects of Generative AI on Collaboration Time}

\begin{figure}[H]
\begin{centering}
\includegraphics[width=1\textwidth]{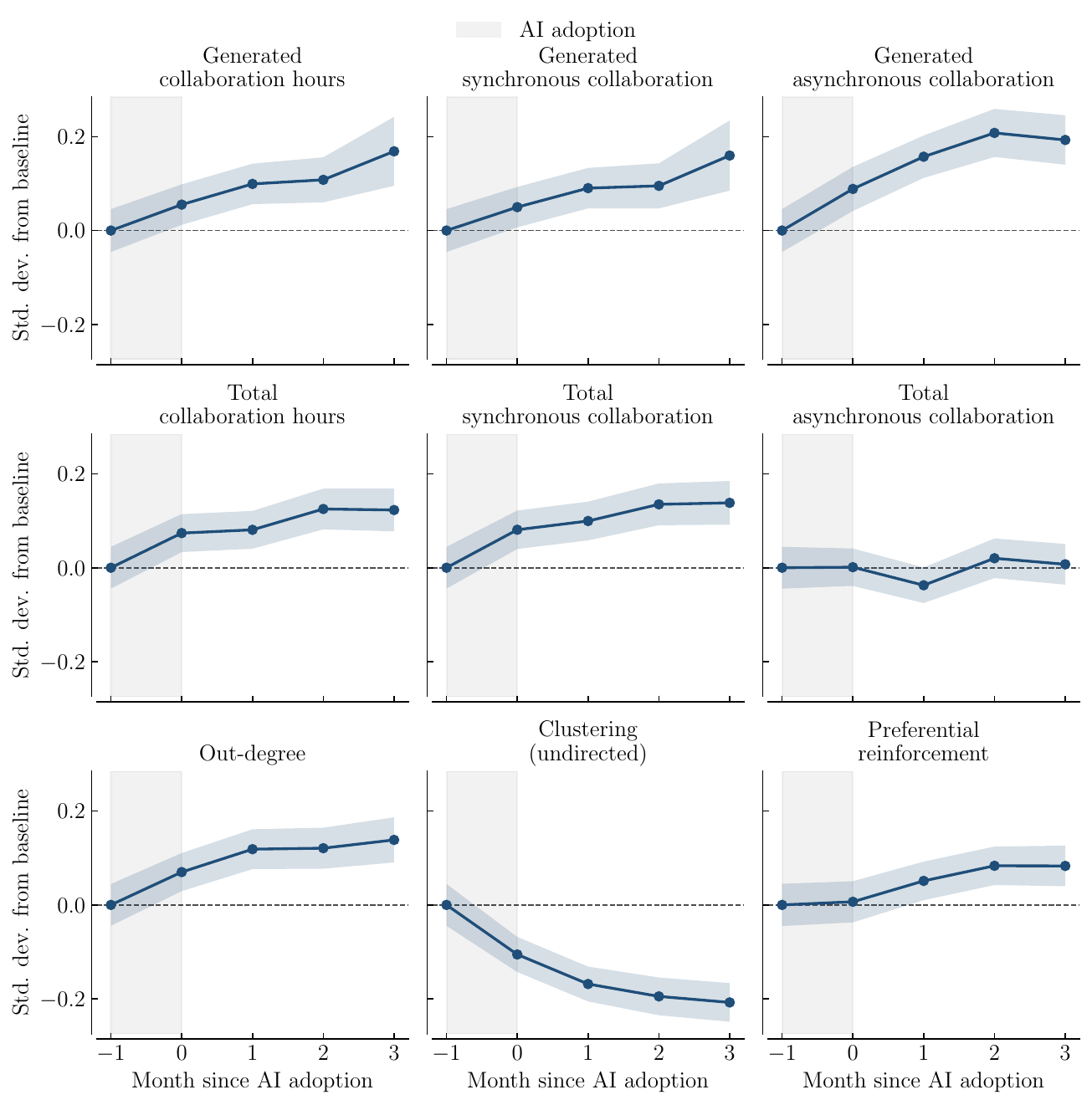}
\par
\caption{Within-individual changes in collaboration and network outcomes around Copilot adoption.}\label{fig:figure1}\end{centering} 
Average outcome trajectories for adopters in event time, centered on each individual's first month of Copilot use ($t = 0$). Each panel reports a different outcome. Values are detrended by calendar-month means (to remove common time shocks), normalized to each individual's pre-adoption baseline at $t = -1$, and expressed in pre-period standard-deviation units to allow visual comparison across outcomes. Solid lines connect monthly averages; shaded bands show 95\% confidence intervals based on the standard error of the mean across individuals. The grey region between $t = -1$ and $t = 0$ marks the month of adoption.
\end{figure}

We begin by examining how Copilot adoption relates to workers' collaboration behavior and network position. Figure \ref{fig:figure1} shows within-individual trajectories in event time for nine outcomes, centered on each worker's first month of Copilot use: generated collaboration hours (top row), total collaboration hours (middle row), and three network measures — out-degree, clustering, and preferential reinforcement (bottom row). Across most outcomes, adopters exhibit visible shifts from their pre-adoption baseline in the months following adoption. 
\begin{center}
\underline{INSERT FIGURE \ref{fig:figure1} ABOUT HERE}
\end{center}

\begin{figure}[H]
\begin{centering}
\includegraphics[width=1\textwidth]{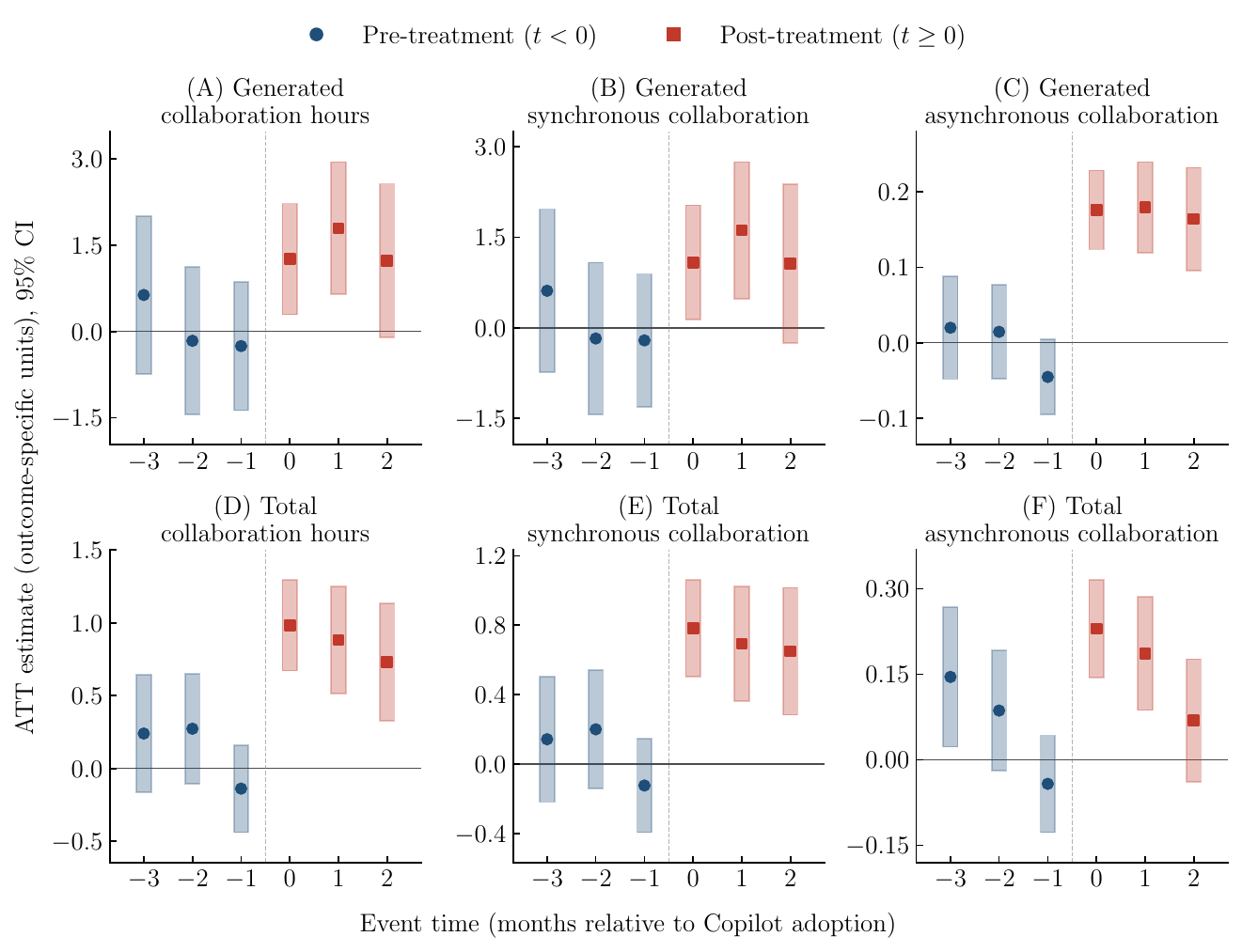}
\par
\centering{}\caption{Event-study estimates of the effect of Copilot adoption on collaboration outcomes (CS-DID).}\label{fig:figure2}\end{centering}
Estimates from the \citet{callaway2021difference} staggered difference-in-differences estimator, reported separately for each outcome over event time $t = -3, \ldots, 2$. Bars show 95\% confidence intervals as filled rectangles bounded by the lower and upper CI limits; markers denote the point estimate. Blue circles correspond to pre-treatment event times ($t < 0$) and red squares to post-treatment event times ($t \geq 0$); the dashed vertical line marks the month of adoption. Pre-treatment coefficients provide a test of the parallel-trends assumption: confidence intervals that contain zero are consistent with parallel trends. Each panel uses its own y-axis (units differ across outcomes).
\end{figure}

Next, we estimate the effect of Copilot adoption using a staggered difference-in-differences design. Figure \ref{fig:figure2} reports the event-study estimates over event time $t = -3, \ldots, 2$, distinguishing outcomes between our primary outcome, generated collaboration time in both asynchronous and synchronous modes (Panels A to C), and total collaboration time (Panels D to F) also articulated in asynchronous and synchronous modes. 
We find no evidence of differential pre-trends prior to adoption. The average pre-treatment effect is close to zero and statistically insignificant ($\beta_{\text{pre\_avg}} = 0.074$; 95\% CI $[-0.962, 1.111]$; $z = 0.14$; $p = 0.889$), while the average post-treatment effect is positive and significant ($\beta_{\text{post\_avg}} = 1.430$; 95\% CI $[0.451, 2.410]$; $z = 2.86$; $p = 0.004$). The event-study coefficients further show that collaboration time rises immediately at adoption ($T_0 = 1.263$; 95\% CI $[0.298, 2.227]$; $z = 2.57$; $p = 0.010$) and remains high one month later ($T_1 = 1.796$; 95\% CI $[0.645, 2.947]$; $z = 3.06$; $p = 0.002$).
We replicated the analysis on the generated load using total collaboration time, which also includes interactions initiated by others. The average pre-treatment effect is close to zero and statistically insignificant ($\beta_{\text{pre\_avg}} = 0.125$; 95\% CI $[-0.176, 0.426]$; $z = 0.81$; $p = 0.417$), while the average post-treatment effect is positive and significant ($\beta_{\text{post\_avg}} = 0.865$; 95\% CI $[0.557, 1.174]$; $z = 5.50$; $p < 0.001$). 

\begin{center}
\underline{INSERT FIGURE \ref{fig:figure2} ABOUT HERE}
\end{center}

\begin{figure}[H]
\begin{centering}
\includegraphics[width=1\textwidth]{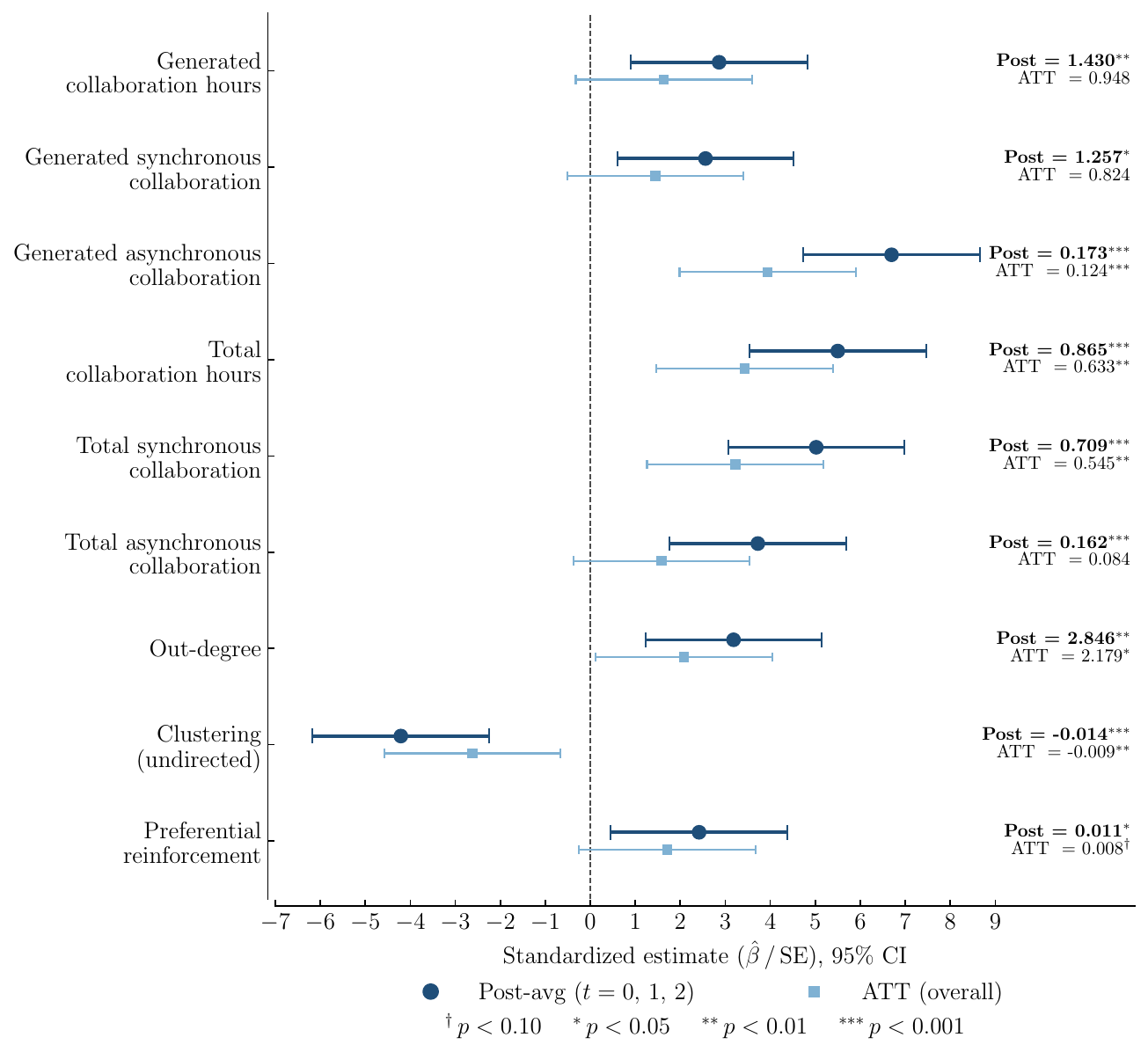}
\par
\caption{Summary treatment effects of Copilot adoption across outcomes (CS-DID).}\label{fig:figure3}\end{centering}
For each outcome, the figure reports two summary measures from the \citet{callaway2021difference} estimator: the Post-avg (dark navy circle), averaging the dynamic event-study coefficients over the first three post-adoption months ($t = 0, 1, 2$), and the overall ATT (light blue square), averaging across all event times and adoption cohorts. Both estimates and their 95\% confidence intervals are standardized by their own standard error ($\beta/\mathrm{SE}$) so that outcomes on different scales share a common x-axis; the dashed vertical line marks zero. Raw coefficients in original units are reported on the right of each row, with significance stars based on each estimator's $p$-value ($^{\dagger}p<0.10$, $^{*}p<0.05$, $^{**}p<0.01$, $^{***}p<0.001$).
\end{figure}

Figure \ref{fig:figure3} shows two estimates for each outcome: the average post-adoption effect (Post-avg), which averages the dynamic event-study coefficients over the first three post-adoption months (Post-avg, $t =0, 1, 2$), and the overall average treatment effect on the treated (ATT), which averages across all event times and adoption cohorts. Coefficients are standardized by their standard errors so that outcomes on different scales can be compared on a common axis; raw coefficients are reported alongside.

For generated collaboration hours, the average effect over the first three post-adoption months is positive and statistically significant (Post-avg $= 1.430$ hours; $z = 2.86$; $p < 0.01$). The overall ATT, which averages across all event times and adoption cohorts, is smaller and not
distinguishable from zero at conventional levels (ATT $= 0.948$; 95\% CI $[-0.187, 2.083]$; $z = 1.64$; $p = 0.102$). The two summaries differ in weighting as well as horizon. The overall ATT averages the group--time effects $ATT(g,t)$ over all cells with $t \geq g$, which mechanically gives more weight to earlier-adopting cohorts because they are observed treated for more periods; the Post-avg instead holds the event window fixed at $e \in \{0,1,2\}$, so every cohort contributes the same three
horizons. If treatment effects vary with adoption timing, the two can differ
through reweighting alone, without any fading of the effect. Because the panel
spans 13 months, only the earliest cohorts are observed at long event times, so
the overall ATT is additionally less precisely estimated at those horizons. We therefore treat the Post-avg as the primary summary and report the full dynamic path in Figure~\ref{fig:figure2}, where any trajectory fading of the initial response can be assessed directly.

We further examine the time spent interacting by collaboration mode, whether synchronous or asynchronous. Specifically, we treat meetings and call hours as synchronous collaboration and email hours as asynchronous collaboration. We excluded chat hours from this decomposition because, unlike meetings, calls, and emails, chat can support either synchronous or asynchronous communication depending on how participants use it, and we lack the behavioral data needed to classify it reliably. We therefore restrict our decomposition to the modes whose classification is unambiguous.\footnote{\url{https://www.microsoft.com/en-us/microsoft-teams/asynchronous-communication}} One interpretation consistent with a validation-driven account is that added coordination would be more likely to emerge in interaction forms suited to rapid clarification and mutual adjustment, that is, in synchronous rather than purely asynchronous exchanges.

\begin{center}
\underline{INSERT FIGURE \ref{fig:figure3} ABOUT HERE}
\end{center}

The event-study estimates show no evidence of differential pre-treatment trends for either mode of collaboration. For synchronous collaboration, the average pre-treatment effect is close to zero and statistically insignificant ($\beta_{\text{pre\_avg}} = 0.078$; 95\% CI $[-0.945, 1.100]$; $z = 0.15$; $p = 0.881$), while the average post-treatment effect within the $[-3, +2]$ window is positive and significant ($\beta_{\text{post\_avg}} = 1.257$; 95\% CI $[0.295, 2.219]$; $z = 2.56$; $p = 0.010$). For asynchronous collaboration, we likewise find no evidence of differential pre-trends ($\beta_{\text{pre\_avg}} = -0.004$; 95\% CI $[-0.054, 0.046]$; $z = -0.14$; $p = 0.885$), and the average post-treatment effect is positive and significant ($\beta_{\text{post\_avg}} = 0.173$; 95\% CI $[0.122, 0.224]$; $z = 6.69$; $p < 0.001$). Thus, Copilot adoption is followed by increases in both synchronous and asynchronous collaboration, with the larger short-run response concentrated in synchronous interaction.

We replicated this analysis using total collaboration time. For synchronous collaboration, the average pre-treatment effect is close to zero and statistically insignificant ($\beta_{\text{pre\_avg}} = 0.073$; 95\% CI $[-0.198, 0.344]$; $z = 0.53$; $p = 0.597$), while the average post-treatment effect is positive and significant ($\beta_{\text{post\_avg}} = 0.709$; 95\% CI $[0.432, 0.986]$; $z = 5.02$; $p < 0.001$). The overall ATT, which averages treatment effects across all treated cohorts and all observed post-adoption periods, is also positive and statistically significant ($\beta_{\text{ATT}} = 0.545$; 95\% CI $[0.214, 0.877]$; $z = 3.22$; $p = 0.001$). For asynchronous collaboration, the average pre-treatment effect is close to zero and statistically insignificant ($\beta_{\text{pre\_avg}} = 0.063$; 95\% CI $[-0.023, 0.149]$; $z = 1.44$; $p = 0.150$), while the average post-treatment effect is positive and significant ($\beta_{\text{post\_avg}} = 0.162$; 95\% CI $[0.077, 0.247]$; $z = 3.72$; $p < 0.001$). The overall ATT is also positive but not statistically significant ($\beta_{\text{ATT}} = 0.084$; 95\% CI $[-0.020, 0.187]$; $z = 1.59$; $p = 0.113$). Taken together, the event-study and ATT estimates suggest that Copilot adoption is followed by an increase in both modes of communication.
Additionally, because collaboration is inherently relational, Copilot adoption may affect not only adopters but also their untreated colleagues, potentially violating the Stable Unit Treatment Value Assumption (SUTVA). In particular, spillovers to untreated employees could contaminate the comparison group and bias our treatment effect estimates. To assess the sensitivity of our findings to this concern, we re-estimate the staggered difference-in-differences models using a restricted sample that excludes never-adopters directly connected to Copilot adopters (network distance = 1), while retaining all treated employees. The estimates on this restricted sample remain consistent with our baseline findings, with larger estimated effects and stronger statistical significance. These results indicate that including untreated employees directly connected to adopters does not change the effect of AI adoption on collaboration.  

\subsection{Effects of Generative AI on Social Networks}

We next examine whether Copilot adoption altered the structure of workers' social networks. If GenAI increases the need for collaboration, then its effects should appear not only in the amount of collaboration workers engage in, but also in the structure of the networks through which that collaboration occurs. As mentioned in section \ref{sec:methods}, we focus on outdegree, local clustering, and strong-tie involvement because together they capture distinct aspects of collaboration that map directly onto our competing predictions.

\begin{figure}[H]
\begin{centering}
\includegraphics[width=1\textwidth]{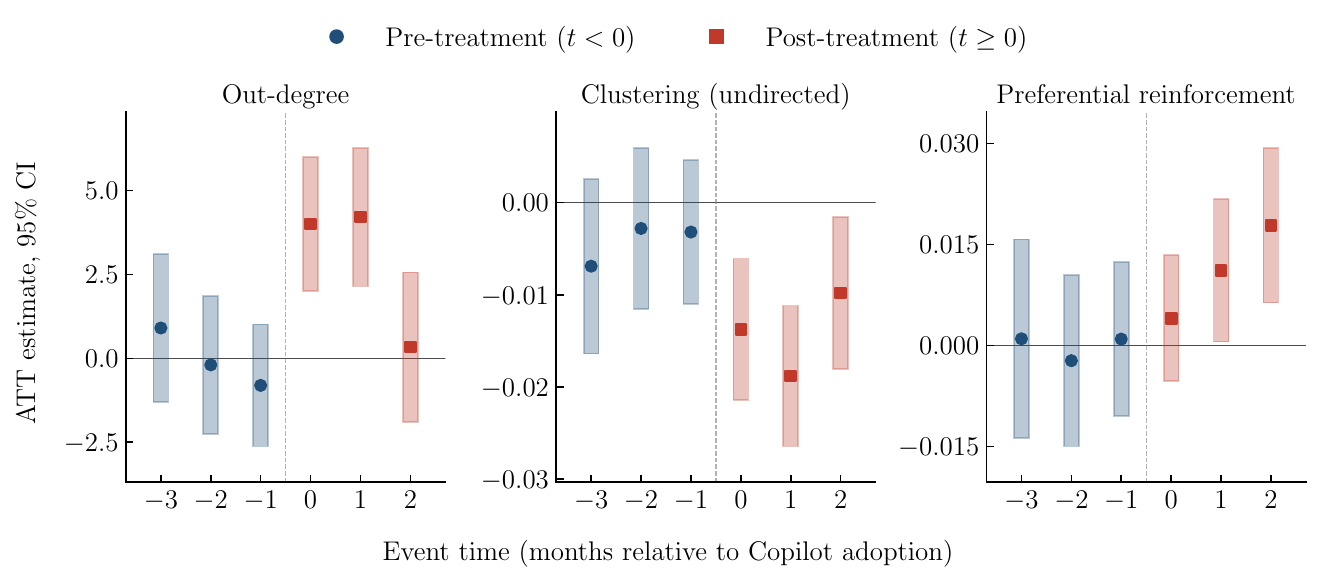}
\par
\centering{}\caption{Event-study estimates of the effect of Copilot adoption on network (CS-DID).}\label{fig:figure4}\end{centering}
Estimates from the \citet{callaway2021difference} staggered difference-in-differences estimator, reported separately for each outcome over event time $t = -3, \ldots, 2$. Bars show 95\% confidence intervals as filled rectangles bounded by the lower and upper CI limits; markers denote the point estimate. Blue circles correspond to pre-treatment event times ($t < 0$) and red squares to post-treatment event times ($t \geq 0$); the dashed vertical line marks the month of adoption. Pre-treatment coefficients provide a test of the parallel-trends assumption: confidence intervals that contain zero are consistent with parallel trends. Each panel uses its own y-axis (units differ across outcomes).
\end{figure}

Figure \ref{fig:figure1} also shows within-individual evolution around Copilot adoption for the key network measures: outdegree, clustering, and strong-tie involvement. We then turn to causal estimates of the effects of Copilot adoption on network structure. Figure \ref{fig:figure4} presents the event-study estimates, while Figure \ref{fig:figure3} shows the corresponding average treatment effects on the treated.

\begin{center}
\underline{INSERT FIGURE \ref{fig:figure4} ABOUT HERE}
\end{center}

The event-study estimates on outdegree show no evidence of differential pre-treatment trends ($\beta_{\text{pre\_avg}} = -0.033$; 95\% CI $[-1.706, 1.642]$; $z = -0.040$; $p = 0.970$). Following adoption, outdegree increases significantly ($\beta_{\text{post\_avg}} = 2.846$; 95\% CI $[1.094, 4.598]$; $z = 3.180$; $p = 0.001$), with the effect emerging immediately after adoption. The overall ATT estimates show that Copilot adoption increases outdegree by approximately 2.2 collaborators ($\beta_{\text{ATT}} = 2.179$; 95\% CI $[0.130, 4.227]$; $z = 2.080$; $p = 0.037$). This indicates that adopters interact with more colleagues after adoption. Rather than narrowing interpersonal exchange, Copilot adoption appears to widen the set of people involved in workers' day-to-day collaboration.

With respect to clustering, the event-study estimates provide no evidence of differential pre-treatment trends ($\beta_{\text{pre\_avg}} = -0.004$; 95\% CI $[-0.012, 0.003]$; $z = -1.150$; $p = 0.248$). After adoption, clustering declines ($\beta_{\text{post\_avg}} = -0.014$; 95\% CI $[-0.021, -0.008]$; $z = -4.210$; $p < 0.001$). The overall ATT is also negative ($\beta_{\text{ATT}} = -0.009$; 95\% CI $[-0.016, -0.002]$; $z = -2.620$; $p = 0.009$), indicating that adopters' networks become less locally closed. Thus, adopters appear to reach beyond dense pockets of collaborators, displaying less redundant collaboration structures.

Finally, for strong ties, the event-study estimates show no evidence of differential pre-treatment trends ($\beta_{\text{pre\_avg}} = -0.000$; 95\% CI $[-0.0108, 0.0106]$; $z = -0.020$; $p = 0.986$). After adoption, reliance on strong ties increases ($\beta_{\text{post\_avg}} = 0.011$; 95\% CI $[0.002, 0.020]$; $z = 2.420$; $p = 0.016$). The overall ATT confirms this pattern ($\beta_{\text{ATT}} = 0.008$; 95\% CI $[-0.001, 0.018]$; $z = 1.71$; $p = 0.088$). This result indicates that network expansion does not come at the expense of trusted or familiar relationships. Instead, Copilot adoption is associated with a dual pattern, such that workers broaden the set of colleagues with whom they interact while also concentrating more of their collaboration on relationships with stronger relational histories.

\subsection{Additional Analysis: TWFE-IV}

Building on the shift-share design described above, we estimate fixed-effects instrumental-variable models of the relationship between Copilot use intensity and collaboration outcomes. The endogenous variable is monthly Copilot use intensity, operationalized as the inverse hyperbolic sine (ihs)\footnote{The ihs transformation is approximately equal to $\log(x)$ for moderate-to-large values of $x$ but, unlike the log, is defined at zero; this is appropriate here because the distribution of Copilot actions is heavily right-skewed and a substantial share of worker-months contain no recorded Copilot activity.} of the total number of Copilot actions taken by worker $i$ in month $t$. The instrument is the interaction between worker-specific baseline exposure to Copilot users and monthly Copilot update intensity. As noted above, baseline exposure is measured in each worker's first observed month in the panel as the share of direct collaborators who were already using Copilot, while update intensity captures Copilot-specific product releases in each month.

The first-stage equation is:
\begin{equation}
X_{i,t} = \pi Z_{i,t} + \alpha_i + \gamma_t + \mu_{i,t}
\end{equation}

where $X_{i,t}$ is Copilot use intensity, $Z_{i,t}$ is the shift-share instrument, $\alpha_i$ are worker fixed effects,  $\gamma_t$ are month fixed effects, and the error
$\mu_{i,t}$ collects idiosyncratic determinants of adoption intensity. 
The second-stage equation is:
\begin{equation}
Y_{i,t} = \beta \widehat{X}_{i,t} + \alpha_i + \gamma_t + \varepsilon_{i,t}
\end{equation}

where $Y_{i,t}$ is the collaboration outcome for worker $i$ in month $t$, and $\beta$ captures the effect of instrumented Copilot use intensity. Standard errors are clustered at the worker level. Worker fixed effects ($\alpha_i$) absorb all time-invariant sources of individual heterogeneity, including baseline collaboration style, role, and personality; month fixed effects ($\gamma_t$) absorb common temporal shocks such as seasonal variation in workload or organization-wide changes in communication norms; $\varepsilon_{i,t}$ collects unobserved determinants of collaboration.

Models are estimated on the same worker-month panel used in the main analyses. We study the same set of outcomes: generated collaboration time, synchronous and asynchronous collaboration time separately, and the three network measures. Table \ref{tab:first_stage} reports the first-stage estimates, while Tables \ref{tab:second_stage_collab} and \ref{tab:second_stage_network} report the second-stage estimates for collaboration time and network outcomes, respectively. The first-stage estimates indicate that the instrument is a strong predictor of monthly Copilot use intensity ($\beta = 0.010$; 95\% CI $[0.008, 0.013]$; $z = 8.97$; $p < 0.001$). The Kleibergen-Paap Wald $F$ statistic is 80.43, well above the Stock-Yogo 10\% critical value of 16.38, indicating a strong first stage \citep{stock2005testing}.

\begin{center}
\underline{INSERT TABLES \ref{tab:first_stage}, \ref{tab:second_stage_collab}, AND \ref{tab:second_stage_network} ABOUT HERE}
\end{center}

\paragraph{Collaboration Time.}
Table~\ref{tab:second_stage_collab} reports the second-stage estimates for
collaboration time. We first focus on collaboration time, both generated by users and total, before moving to the analysis of synchronous and asynchronous collaboration modes. Model~1 estimates the effect of instrumented Copilot use
intensity on \emph{user-initiated} collaboration hours: the coefficient is
positive but only marginally significant ($\beta = 1.552$; 95\% CI
$[-0.216,\,3.320]$; $z = 1.72$; $p = 0.085$). Model~4 estimates the same
specification for \emph{total} collaboration time and yields a positive and
highly significant effect ($\beta = 1.593$; 95\% CI $[0.795,\,2.391]$;
$z = 3.91$; $p < 0.001$). The two point estimates are almost identical in magnitude. Because the
regressor enters in inverse hyperbolic sine form, a tenfold increase in
monthly Copilot actions---for instance, from 1 to 10 actions per month---
corresponds to a change of $\beta \times [\operatorname{asinh}(10) - \operatorname{asinh}(1)]$, that is, about $3.3$ additional user-initiated collaboration hours per month in Model~1 and about $3.4$ additional total collaboration hours in Model~4.

Disaggregating the results by channel mode, Model 2 shows that the effect of instrumented Copilot intensity on user-generated synchronous collaboration hours is positive, as in the main difference-in-differences analysis, but it is not significant at the conventional level ($\beta = 1.329$; 95\% CI $[-0.417, 3.06]$; $z = 1.49$; $p = 0.136$). Model 5 shows the effect on total synchronous collaboration, which is positive and statistically significant ($\beta = 1.233$; 95\% CI $[0.549, 1.919]$; $z = 3.53$; $p < 0.001$). For asynchronous mode, Model 3 shows a positive and statistically significant effect of Copilot intensity on user-generated collaboration hours ($\beta = 0.222$; 95\% CI $[0.123, 0.322]$; $z = 4.38$; $p < 0.001$). And Model 6 replicates this finding on total asynchronous collaboration time ($\beta = 0.417$; 95\% CI $[0.195, 0.639]$; $z = 3.68$; $p < 0.001$).

\paragraph{Social Networks.} Table \ref{tab:second_stage_network} reports the network-level consequences of GenAI use. Model 1 shows that Copilot use intensity increases outdegree ($\beta = 9.396$; 95\% CI $[5.554, 13.237]$; $t = 4.79$; $p < 0.001$). Each tenfold increase in Copilot actions leads to reaching approximately 20 additional distinct colleagues per month ($9.396 \times [\operatorname{asinh}(10) - \operatorname{asinh}(1)]$). Model 2 examines clustering, displaying that Copilot use intensity decreases local clustering coefficient ($\beta = -0.030$; 95\% CI $[-0.056, -0.004]$; $t = -2.23$; $p = 0.026$). Therefore, more intensive Copilot use yields less redundancy in workers' network, as they reach beyond tightly interconnected local circles. Model 3 focuses on strong ties defined as recurrent, high-intensity collaborative relationships. Here, greater Copilot use intensity leads workers to allocate more time to contacts with whom they had interacted more extensively in the past ($\beta = 0.031$; 95\% CI $[0.005,\,0.058]$; $t = 2.30$; $p = 0.021$). This result indicates that more intensive GenAI use not only expands workers' reach but also deepens a subset of existing relationships. 
Taken together, the network results extend the collaboration-time findings. More intensive GenAI use is associated with networks that are broader and less clustered, but also relationally deeper: workers reach outward for diverse expertise while leaning more heavily on a stable core of colleagues for
validation and judgment.

\section{Discussion}
A widely held belief is that GenAI will increase the independence of knowledge workers by reducing the need for human-to-human collaboration in everyday activities \citep{roulet2025overreliance}. This increased independence may isolate individuals from their social context, echoing patterns observed in remote work \citep{yang2022effects}. Challenging this view, we find that generative AI makes knowledge work more social: those who use it collaborate more frequently, draw upon a broader group of colleagues, move beyond the small groups they were originally part of, and place greater reliance on the few individuals they trust the most. The technology appears to reorganize the social structure of work rather than make it more transient.

One interpretation consistent with these patterns is that GenAI changes the relative costs of production and coordination. Although these new tools make it easier for workers to generate outputs such as drafts, summaries, and recommendations, such outputs are probabilistic and opaque by nature, making them difficult to interpret and verify \citep{lebovitz2022engage}. When outputs are context-sensitive, their fluency can outpace users' ability to assess accuracy, detect subtle errors, or anticipate downstream implications. As a result, the key challenge in knowledge work may shift from generating correct answers to securing their social acceptance, a responsibility that relies on human-to-human interactions \citep{anthony2021knowledge}.

In these circumstances, workers may consult colleagues not only to co-produce but also to align interpretations. Because AI-assisted outputs require social validation, knowledge workers devote more time to collaboration and engage more broadly across the organization to obtain context and independent judgment \citep{friedkin1990social}.

Collaboration networks become more expansive and less clustered as workers reach outward precisely because generative AI has raised, rather than lowered, the stakes of social acceptance. At the same time, broader social reach does not imply a uniform expansion of collaboration. Workers who use AI appear to interact more heavily with a subset of already-trusted colleagues. These patterns suggest that workers may be searching for expertise and independent judgment. Thus, interpersonal interactions act as a system for checking and validating AI-assisted outputs, using social processes to decide whether to accept, change, or reject them.

These findings add a relational dimension to a literature that has focused primarily on productivity and cognition \citep{brynjolfsson2025generative,dellacqua2026navigating}. Prior research has shown that GenAI can substantially improve both efficiency and output quality in tasks such as professional writing and customer support \citep{shin2026adoption}, especially benefiting less-experienced workers \citep{noy2023experimental}. In this study, we show that adoption of GenAI contributes to reshaping workers' social structure in ways that differ from other major technology-enabled transitions \citep{barley1990alignment}.

More generally, this finding suggests that the social effects of GenAI could be systematically misjudged if they are mainly seen in terms of substituting human-to-human collaboration, making workers more isolated. This increased collaboration reveals that knowledge workers' trust towards AI-assisted outputs is contingent upon engaging in parallel social processes such as consultation, deliberation, and peer review. 

Thus, social relationships among colleagues can compensate for the opacity of GenAI. The concept of epistemic debt qualifies this interpretation \citep{unceta2026epistemic}. Collaboration pays down the debt only when colleagues interrogate an output, asking for its reasoning and testing its assumptions. Collaboration can also spread the epistemic debt. An output endorsed by a trusted colleague feels settled, others may build on it without examining it, and unexamined claims gain authority as they circulate. This distinction maps onto our two network findings. Broader, less clustered networks bring in evaluators whose knowledge overlaps less with the worker’s own, which is the condition for independent scrutiny. Greater reliance on strong ties is harder to read: trusted colleagues offer informed judgment, but trust also lowers the scrutiny people apply to each other’s work. Our data cannot separate these functions, because we observe who interacts with whom and for how long, not what they discuss. Distinguishing validation from endorsement is therefore a priority for future research. It also matters for practice. Organizations that want collaboration to offset GenAI's opacity need norms that reward questioning AI-assisted work and disclosing when it was used.

Moreover, such a compensatory mechanism can mitigate the risk of low-quality AI-generated outputs. As research on Co-Scientist---an AI-assisted scientific discovery---demonstrates \citep{gottweis2026coscientist}, although AI systems can support hypothesis generation and refinement, they also require appropriate safeguards and oversight to ensure that the quality of their outputs is not compromised. Hence, extensive collaborative structure can provide the social scaffolding necessary for these cross checks.

An open question for future research is how these individual-level shifts in collaboration behavior scale up to the network's overall structure. The aggregate patterns suggest that the effect goes beyond a simple increase in activity, triggering a systematic reorganization of how connections are distributed. Additional network-level analysis in our research context reveals that, as the network grows, edges increase faster than nodes, leading to important structural changes: path lengths and diameter decline, global efficiency rises, and clustering does not keep pace with degree. The resulting network is therefore more integrated than locally cohesive as distances between actors shrink.

An important consideration is determining when the collaborative intensification associated with generative AI enhances judgment and when it introduces friction or cognitive overload. Future research should assess how these effects differ across tasks that vary in verifiability, ambiguity, accountability, and interdependence, as these dimensions are likely to influence both the demand for collective evaluation and the associated costs. Further investigation is needed to identify systems that allow firms to satisfy validation requirements without imposing excessive coordination burdens. Potential solutions include establishing clearer protocols for reviewing AI-assisted outputs, explicitly assigning accountability for AI-supported decisions, and developing specialized validation roles.

There are several important limitations to our study. First, we focus on a single business unit within a large multinational company, so it is not clear if our findings apply to other industries, workplaces, job types, or institutional settings. For instance, in jobs where workers do not depend much on each other, like software development, GenAI might actually reduce informal collaboration \citep{ulfsnes2024transforming}. Second, our main explanation—that workers consult more widely because it is hard to judge generative outputs—is based on behavioral data, not direct observation. We know who interacts with whom and for how long, but not what they talk about. We also do not measure how uncertain workers feel about AI-generated content or how much they need to check it. Other explanations could also fit the data: people might use time saved from drafting to interact more, some output could require more coordination even if no one doubts the quality, and some discussions might be about the technology itself rather than its results. Moreover, our findings cover only the first three months after adoption, which is when collaboration patterns are changing. We do not know if these patterns will stay the same over time; longer studies are needed to answer that. To tell these explanations apart, we would need data on what people discuss or ways to measure how uncertain they feel. Third, while our event study with staggered difference-in-differences and an instrumental variable approach helps reduce concerns about confounding from earlier trends, our causal claims still depend on the validity of our main assumptions.


\clearpage
\pagebreak
{
\singlespacing
\bibliographystyle{apalike}
\bibliography{bibliography}
}
\pagebreak
\appendix
\section*{Tables and Figures}


\begin{table}[H]
\centering
\caption{First-stage regression results}
\label{tab:first_stage}
\begin{threeparttable}
\footnotesize
\begin{tabular}{lc}
\toprule
 & Copilot actions\\
\midrule
Instrument
    & $0.010^{***}$ \\
    & $(0.008,\,0.013)$ \\
    & $t = 8.97$ \\
    & $p < 0.001$ \\
\addlinespace
Month FE        & Yes \\
User FE         & Yes \\
Observations    & 141{,}787 \\
Users           & 11{,}374 \\
$R^2$ (within)  & 0.092 \\
\bottomrule
\end{tabular}
\begin{tablenotes}[flushleft]\footnotesize
\item \textit{Notes.} First-stage fixed-effects regression from the TWFE-IV specification. The dependent variable is the inverse hyperbolic sine of total Copilot actions. The unit of observation is the individual--month $(i,t)$. The specification includes user and month fixed effects. The values in parentheses below each coefficient are 95\% confidence intervals computed from robust standard errors clustered at the user level. All $p$ values are two-sided.
\item \textit{Significance:} $^{\dagger}p<0.10$, $^{*}p<0.05$, $^{**}p<0.01$, $^{***}p<0.001$.
\end{tablenotes}
\end{threeparttable}
\end{table}

\begin{table}[H]
\centering
\caption{Second-stage IV regression results: effect of Copilot actions on collaboration outcomes}
\label{tab:second_stage_collab}
\begin{threeparttable}
\setlength{\tabcolsep}{5pt}
\footnotesize
\begin{tabular}{lcccccc}
\toprule
 & \multicolumn{3}{c}{Generated Collaboration (Hours)} & \multicolumn{3}{c}{Total Collaboration (Hours)} \\
\cmidrule(lr){2-4} \cmidrule(lr){5-7}
 & All Modes & Sync. & Async. & All Modes & Sync. & Async. \\
 & (1) & (2) & (3) & (4) & (5) & (6) \\
\midrule
Copilot actions
    & $1.552^{\dagger}$
    & $1.329$
    & $0.223^{***}$
    & $1.593^{***}$
    & $1.234^{***}$
    & $0.417^{***}$ \\
    & $(-0.22,\,3.32)$
    & $(-0.42,\,3.08)$
    & $(0.12,\,0.32)$
    & $(0.79,\,2.39)$
    & $(0.55,\,1.92)$
    & $(0.20,\,0.64)$ \\
    & $z = 1.72$
    & $z = 1.49$
    & $z = 4.38$
    & $z = 3.91$
    & $z = 3.53$
    & $z = 3.68$ \\
    & $p = 0.085$
    & $p = 0.136$
    & $p < 0.001$
    & $p < 0.001$
    & $p < 0.001$
    & $p < 0.001$ \\
\addlinespace
Month FE        & Yes & Yes & Yes & Yes & Yes & Yes \\
User FE         & Yes & Yes & Yes & Yes & Yes & Yes \\
Observations    & 141{,}787 & 141{,}787 & 141{,}787 & 149{,}945 & 149{,}945 & 149{,}945 \\
Users           & 11{,}374 & 11{,}374 & 11{,}374 & 12{,}030 & 12{,}030 & 12{,}030 \\
$R^2$ (within)  & 0.011 & 0.009 & 0.030 & 0.090 & 0.098 & 0.045 \\
\bottomrule
\end{tabular}
\begin{tablenotes}[flushleft]\footnotesize
\item \textit{Notes.} Second-stage fixed-effects IV estimates from the TWFE specification. The endogenous regressor, Copilot actions (ihs), is instrumented using the shift-share instrument. Generated collaboration refers to ego-initiated communication; Total collaboration includes both incoming and outgoing collaboration time. Each column reports a separate regression. The unit of observation is the individual--month $(i,t)$. All specifications include user and month fixed effects. The values in parentheses below each coefficient are 95\% confidence intervals computed from robust standard errors clustered at the user level. All $p$ values are two-sided.
\item \textit{Significance:} $^{\dagger}p<0.10$, $^{*}p<0.05$, $^{**}p<0.01$, $^{***}p<0.001$.
\end{tablenotes}
\end{threeparttable}
\end{table}

\begin{table}[H]
\centering
\caption{Second-stage IV regression results: effect of Copilot actions on network outcomes}
\label{tab:second_stage_network}
\begin{threeparttable}
\footnotesize
\begin{tabular}{lccc}
\toprule
 & Out-degree & Clustering & Pref. reinforcement \\
 & (1) & (2) & (3) \\
\midrule
Copilot actions
    & $9.396^{***}$
    & $-0.030^{*}$
    & $0.031^{*}$ \\
    & $(5.55,\,13.24)$
    & $(-0.056,\,-0.004)$
    & $(0.005,\,0.058)$ \\
    & $z = 4.79$
    & $z = -2.23$
    & $z = 2.30$ \\
    & $p < 0.001$
    & $p = 0.026$
    & $p = 0.021$ \\
\addlinespace
Month FE        & Yes & Yes & Yes \\
User FE         & Yes & Yes & Yes \\
Observations    & 149{,}945 & 149{,}945 & 135{,}252 \\
Users           & 12{,}030 & 12{,}030 & 11{,}995 \\
$R^2$ (within)  & 0.129 & 0.010 & 0.012 \\
\bottomrule
\end{tabular}
\begin{tablenotes}[flushleft]\footnotesize
\item \textit{Notes.} Second-stage fixed-effects IV estimates from the TWFE specification. The endogenous regressor, Copilot actions (ihs), is instrumented using the shift-share instrument. Out-degree counts the number of distinct collaborators per month; clustering is measured on the undirected network; preferential reinforcement captures the tendency to repeatedly interact with the same collaborators. Each column reports a separate regression. The unit of observation is the individual--month $(i,t)$. All specifications include user and month fixed effects. The values in parentheses below each coefficient are 95\% confidence intervals computed from robust standard errors clustered at the user level. All $p$ values are two-sided.
\item \textit{Significance:} $^{\dagger}p<0.10$, $^{*}p<0.05$, $^{**}p<0.01$, $^{***}p<0.001$.
\end{tablenotes}
\end{threeparttable}
\end{table}

\end{document}